\documentclass{elsart}

\usepackage{amstext}
\usepackage{graphicx}

\makeatletter

\def\ps@copyright{\let\@mkboth\@gobbletwo
  \def\@oddhead{}\def\@evenhead{}
  \def\@oddfoot{}\def\@evenfoot{}}

\makeatother

\newcommand\stext[1]{\text{\scriptsize#1}}
\newcommand\MeV[1]{#1~\text{MeV}}

\begin{document}

\begin{frontmatter}

\title{On the linear term in the nuclear symmetry energy}

\author{Kai Neerg{\aa}rd}

\address{N{\ae}stved Gymnasium og HF,
  Nyg{\aa}rdsvej 43, DK-4700 N{\ae}stved, Denmark}

\ead{neergard@inet.uni2.dk}

\begin{abstract}

The nuclear symmetry energy calculated in the RPA from the pairing plus
symmetry force Hamiltonian with equidistant single-nucleon levels is for
mass number $A=48$ approximately proportional to $T(T+1.025)$, where $T$
is the isospin quantum number. An isovector character of the pair field
assumed to produce the observed odd-even mass staggering is essential
for this result. The RPA contribution to the symmetry energy cannot be
simulated by adding to the Hartree-Fock-Bogolyubov energy a term
proportional to the isospin variance in the Bogolyubov quasiparticle
vacuum, and there are significant corrections to the approximation which
consist in adding half the isocranking angular velocity. The present
calculation employs a smaller single-nucleon level spacing than used in
a previous investigation of the model.

\end{abstract}


\end{frontmatter}

The so-called symmetry energy in the semi-empirical formula for nuclear
masses is quadratic in the difference $N-Z$ between the numbers of
neutrons and protons~\cite{Wei}. In a more general view including
isobaric analogue states, this may be seen as a quadratic dependence on
the isospin quantum number $T$. Nuclei with small values of $T$ have a
larger binding energy than expected from a quadratic extrapolation from
higher $T$-values~\cite{MySw66}. In one parametrization of this extra
binding energy, the so-called Wigner energy, a term linear in $T$ is
included in the total energy so that the symmetry energy becomes
proportional to $T(T+x)$ for some constant $x$. The best choice of $x$
seems to be close to 1~\cite{Jae,Mach}, possibly with a preference for a
slightly larger value~\cite{Gl}. Other parametrizations of the Wigner
energy are employed in Refs.~\cite{MySw66} and~\cite{Gor}.
 
The origin of the Wigner energy is a debated
issue~\cite{Wi,Ta,MySw66,MySw97,SaWy97,Ci,Sa,Po,SaWy00,Roe,Gor}. For a
schematic, isobarically invariant Hamiltonian, I recently showed that
the Hartree-Bogolyubov energy rises quadratically in $T$, whereas the
$T$-dependence of the RPA correlation energy is dominated by a term
equal to half the derivative of the Hartree-Bogolyubov energy with
respect to $T$. The Hartree-Bogolyubov energy and the dominant
$T$-dependent term in the RPA correlation energy thus give a symmetry
energy proportional to $T(T+1)$~\cite{Ne}.

For the details of the theory, see Ref.~\cite{Ne}. The Hamiltonian is
the sum of an independent-nucleon Hamiltonian with equidistant, fourfold
degenerate single-nucleon levels, an isobarically invariant isovector
pairing force and a symmetry force:
\[
		H = H_0 - G\mathbf P^\dagger\cdot\mathbf P
    + \frac{\kappa{\mathbf T^2}}2\,,\quad
		H_0 = \sum_{k\sigma\tau}\epsilon_k
		a^\dagger_{k\sigma\tau}a^{\phantom\dagger}_{k\sigma\tau}\,.
\]
Here, $\mathbf P$ is a pair annihilation isovector, $\mathbf T$ is the
isospin, and $G$ and $\kappa$ are coupling constants. The single-nucleon
energy $\epsilon_k$ takes $\Omega$ equidistant values separated by
$\eta$, and $a_{k\sigma\tau}$ are single-nucleon annihilation operators.
$k\sigma$ is $k$ or $\overline k$, where the bar denotes time reversal,
and $\tau$ distinguishes neutrons from protons. Hartree-Bogolyubov
optimal states are derived from a Routhian
\[
  R = H - \lambda A_{\stext v} - \mu T_z\,,
\]
where $A_{\stext v}$ is the number of valence nucleons. The nucleon
chemical potential $\lambda$ is placed midway between the lowest and
the highest $\epsilon_k$, so that $\langle A_{\stext v}\rangle=2\Omega$
in the optimal state, and the isocranking angular velocity $\mu$ varied
to produce a range of $\langle T_z\rangle$.

The symmetry of even $A_{\stext v}/2\pm T_z$ is imposed on the trial
Bogolyubov quasiparticle vacuum. The model thus describes isobaric
multiplets whose substate with $T_z=T$ belongs to a doubly even nucleus.
Due to this symmetry, the $z$-coordinate of the gap isovector
$\mathbf\Delta=-G\langle\mathbf P\rangle$ vanishes. By a suitable choice
of phases, the optimal state is then an ordinary product of neutron and
proton BCS states with a common real neutron and proton gap $\Delta$.
For numeric convenience, $\Delta$ is kept constant with the variation of
$\mu$, which results in a negligible variation of $G$.

The RPA calculation employs the standard technique of a perturbative
boson expansion of products of nucleon field operators~\cite{Mar}. By
truncating the resulting expansions of $H$ and $R$ to second order, RPA
operators $\tilde H$ and $\tilde R$ are obtained. The initial symmetries
of $H$ and $R$ are recovered by replacing in $\tilde H$ and $\tilde R$
truncated expansions of $A_{\stext v}$ and $\mathbf T$ by formal
variables which obey the mutual commutation relations of the exact
operators and have the same commutators with other boson operators as
the truncated ones. When $\langle A_{\stext v}\rangle$ and $\langle
T_z\rangle$ are equal to eigenvalues of $A_{\stext v}$ and $T_z$
respectively, the lowest eigenstate of the approximate Routhian thus
obtained has $A_{\stext v}=\langle A_{\stext v}\rangle$ and
$T=T_z=\langle T_z\rangle$. This state is also an eigenstate of the
approximate Hamiltonian with an eigenvalue $E$ that is the sum of two
terms: the Hartree-Bogolyubov energy
\[
  E_0 = \langle H_0\rangle - \frac{2\Delta^2}G + \frac{\kappa T^2}2
\]
and the RPA correlation energy
\begin{equation}\label{E2-1}
  E_2 = \left\langle - G\Delta\mathbf P^\dagger\cdot\Delta\mathbf P
    + \frac{\kappa\Delta\mathbf T^2}2\right\rangle
    + {\textstyle\frac12}\left(\sum_\nu\omega_\nu
      - \sum_{i<j}[b^{\phantom\dagger}_{ij},[\tilde R,b^\dagger_{ij}]]
      \right)\,.
\end{equation}
Here, $\Delta\mathbf P=\mathbf P-\langle\mathbf P\rangle$ and
$\Delta\mathbf T=\mathbf T-\langle\mathbf T\rangle$, $\omega_\nu$
denotes the eigenfrequencies of the boson system with Hamiltonian
$\tilde R$, and $b^{\phantom\dagger}_{ij}$ are the boson annihilation
operators associated with pairs of Bogolyubov quasiparticles. $E_0$,
$E_2$ and $E=E_0+E_2$ are defined for any $T=\langle T_z\rangle$ by
these expressions.

In the calculation of Ref.~\cite{Ne}, which was intended to simulate the
isobaric chain with mass number $A=48$, I used the Fermi gas estimate of
the single-nucleon level spacing $\eta$. As pointed out by Satula and
Wyss~\cite{SaWy02} and G\l owacz, Satula and Wyss~\cite{Gl}, this gives
an unrealistically large value. I have therefore repeated the
calculation for two smaller values: $\eta=4/(6A/(\MeV{c})/\pi^2)$ with
$c=8$ and 10~\cite{GiCa,BoMo,Ka}, which gives $\eta=1.1$ and \MeV{1.4}
respectively. Otherwise, the parameters of the present calculation are
the same as in Ref.~\cite{Ne}: The number $\Omega$ of equidistant,
fourfold degenerate single-nucleon levels is equal to 24, the pair gap
$\Delta=\MeV{12}/\sqrt{A}=\MeV{1.7}$~\cite{BoMo}, and the total
coefficient of $T^2/2$ in the Hartree-Bogolyubov part of the symmetry
energy
$\eta+\kappa=2\MeV{(134.4A^{-1}-203.6A^{-4/3})}=\MeV{3.3}$~\cite{DuZu}.

\begin{figure}
{\center\includegraphics[width=.8\textwidth]{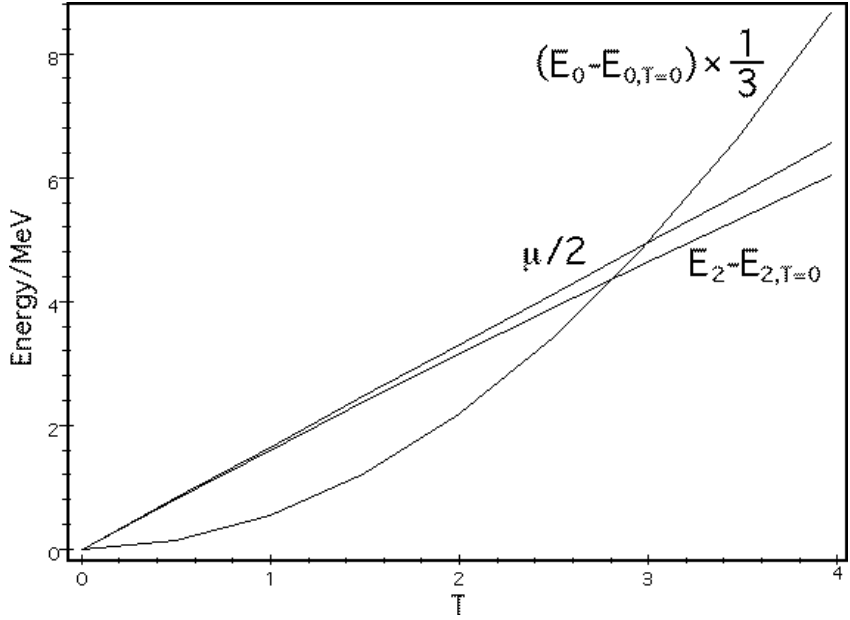}\par}
\caption[]{\label{fig1}$E_0-E_{0,T=0}$, $E_2-E_{2,T=0}$ and $\mu/2$ as
functions of $T$ for $\eta=\MeV{1.1}$. $E_0-E_{0,T=0}$ is scaled by a
factor $1/3$.}
\end{figure}

The result for the smallest single-nucleon level spacing
$\eta=\MeV{1.1}$ is shown in Fig.~\ref{fig1}.
A comparison with Fig.~1 of Ref.~\cite{Ne} shows that the change
of $\eta$ does not alter the situation qualitatively.%
\footnote{Due to a computation error, $\mu/2-(E_2-E_{2,T=0})$ as shown
in Fig.~1 of Ref.~\cite{Ne} is too large by an almost constant factor
about 1.2.}
Still, almost exactly $E_0-E_{0,T=0}=(\eta+\kappa)T^2/2$. Since the
pairing energy $2\Delta^2/G$ is essentially constant---it varies by less
than \MeV{.1} in the range of $T$ considered---this means that even in
the presence of pairing, the `kinetic' contribution to the symmetry
energy, that is, the contribution from $\langle H_0\rangle$, is in a
very good approximation equal to $\eta T^2/2$. This expression is easily
shown to be exact for $\Delta=0$.

Also like in the previous calculation, the RPA contribution to the
symmetry energy $E_2-E_{2,T=0}$ is dominated by the frequency $\mu$ of
the normal mode associated with the direction of the isospin. This is
indeed more accurately true here than for larger values of $\eta$. Thus,
from $T=0$ to 4 the sum of all the other terms in $E_2$ only change by
about \MeV{.5} for $\eta=\MeV{1.1}$ and about \MeV{.8} for
$\eta=\MeV{1.4}$ as compared to about \MeV{1.5} for the value
$\eta=\MeV{2.1}$ given by the Fermi gas estimate. The reason for this
seems obvious: With the smaller level spacing $\eta$, the isodeformation
$\Delta/\eta$ is larger and the isorotation therefore more collective.
This weakens the coupling between the collective and intrinsic degrees
of freedom. It may be noticed that since $\Delta\propto A^{-1/2}$ and
$\eta\propto A^{-1}$, the isodeformation increases with $A$. Heavier
nuclei should therefore show an even more collective isorotation.

The identification of $T$ with $\langle T_z\rangle$ provides an
extension of $E_2(T)$ to negative $T$. Due to the isobaric invariance of
$H$, this results in an even function. The only term in $E_2$ which is
not analytic at $T=0$ is $\mu/2$, which is replaced by $-\mu/2$ for
negative $T$. It follows that $E_2-E_{2,T=0}-\mu/2$ vanishes
quadratically for $T\rightarrow0^+$. Now, apart from a negligible
correction due to the fact that $\Delta$ rather than $G$ is kept
constant in the calculation, we have $\mu=dE_0/dT=(\eta+\kappa)T$. The
curve in Fig.~\ref{fig1} is in fact well approximated by the second
order polynomial $E_2-E_{2,T=0}=(\eta+\kappa)T/2-\MeV{.033}\times T^2$.
Altogether, we thus have in a good approximation
\[
  E-E_{T=0} = \left(\frac{\eta+\kappa}2-\MeV{.033}\right)T(T+x)\,,
\]
where
\[
  x=\left(1-\frac{\MeV{.033}}{(\eta+\kappa)/2}\right)^{-1}=1.020\,.
\]
For $\eta=\MeV{1.4}$, this analysis gives $x=1.031$. These values are
consistent with the possible indications in the data of $x$ being
slightly larger than 1~\cite{Gl}.

From my results in Ref.~\cite{Ne}, G\l owacz, Satula and Wyss infer
$x\approx.8$ for $T=2$~\cite{Gl}. These authors may have calculated
$(E_2-E_{2,T=0})T/(E_0-E_{0,T=0})$, which is evidently not a right way
of extracting $x$ from the energies.

Nuclear masses are extensively compared with Hartree-Fock-Bogolyubov
calculations. See for example Ref.~\cite{Gor} and references therein. If
we neglect the change of optimal state by the exchange terms in the
self-consistent single-nucleon potential and pair field, the difference
between the Hartree-Fock-Bogolyubov and Hartree-Bogolyubov energies, the
Fock term, is in the present theory precisely the first term in
Expr.~(\ref{E2-1}). This is a somewhat artificial term since it is
cancelled by a part of the second sum in the second term.%
\footnote{In Ref.~\cite{Ne}, in the third paragraph of the second colomn
on page 289, `second term in the second sum' should be as here `second
sum in the second term'. It is understood there that if $\Delta=0$, also
$G=0$.}
In fact Eq.~(\ref{E2-1}) may be recast into the form
\begin{equation}\label{E2-2}
  E_2 = {\textstyle\frac12}
    \left(\sum_\nu\omega_\nu - \sum_{i<j}(E_i+E_j)\right)\,,
\end{equation}
where $E_i$ is the Bogolyubov quasiparticle energy.

\begin{figure}
{\center\includegraphics[width=.8\textwidth]{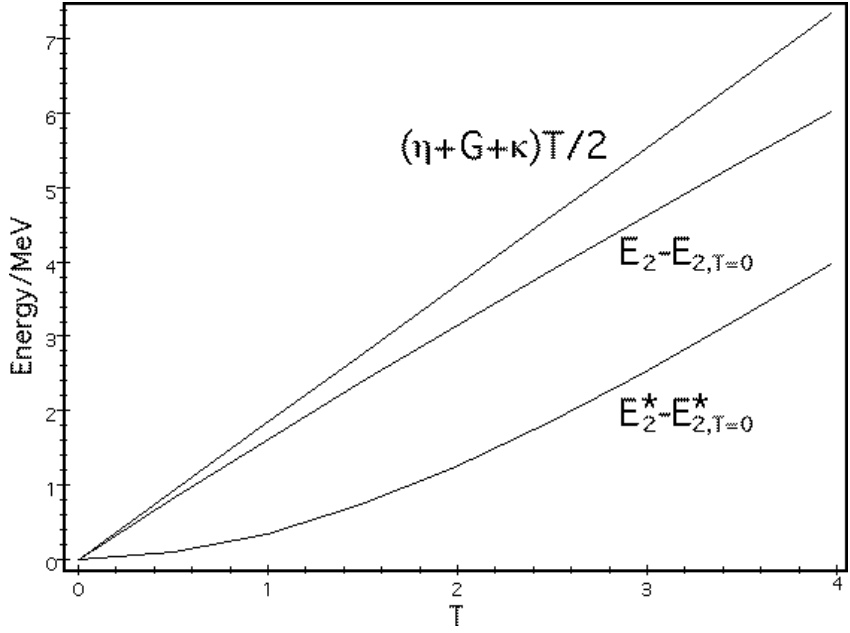}\par}
\caption[]{\label{fig2}$(\eta+G+\kappa)T/2$, $E_2-E_{2,T=0}$ and
$E^\ast_2-E^\ast_{2,T=0}$ as functions of $T$ for $\eta=\MeV{1.1}$.}
\end{figure}

Satula and Wyss seem to suggest that the RPA correlation energy may be
simulated by adding $\eta\langle\Delta\mathbf T^2\rangle/2$ to the
Hartree-Fock-Bogolyubov energy~\cite{SaWy02}. Neglecting again the
change of optimal state, this amounts in the present theory to replacing
$E_2$ by
\[
  E^\ast_2 = \left\langle - G\Delta\mathbf P^\dagger\cdot\Delta\mathbf P
    + \frac{(\eta+\kappa)\Delta\mathbf T^2}2\right\rangle\,.
\]
Since this is a modification of the Fock term, which is absent in
Eq.~(\ref{E2-2}), a relationship between $E_2$ and $E^\ast_2$ is not
obviously anticipated. Fig.~\ref{fig2}
shows that in fact they depend on $T$ very differently. In particular,
$E^\ast_2(T)$ is even and analytic at $T=0$, so
$E^\ast_2-E^\ast_{2,T=0}$ vanishes quadratically for $T\rightarrow0$.
Thus $E^\ast_2$ does not give the empirical cusp at $T=0$ in the graph
of the function $E(T)$.

The $T$-dependence of $E^\ast_2$ mainly reflects that of the isospin
variance $\langle\Delta\mathbf T^2\rangle$. For $\Delta=0$, we have
$\langle\Delta\mathbf T^2\rangle=T$. For $\Delta>0$, the diffuseness of
the nucleon numbers makes a contribution to $\langle\Delta\mathbf
T^2\rangle$ which is largest for $T=0$. $\langle\Delta\mathbf
T^2\rangle$ is an even function of $T$ and for $\Delta>0$ analytic at
$T=0$. This results in the hyperbola-like behaviour seen in the inset of
Fig. 3 of Ref.~\cite{SaWy02} and reflected in Fig.~\ref{fig2}. For
$\Delta=0$, the first term in $E^\ast_2$ has a $T$-dependent part
$GT/2$, which arises from blocking of the neutron-proton pairing force
when nucleons are promoted from proton to neutron levels. For
$\Delta>0$, the asymptotic slope of the hyperbola gets a corresponding
contribution from this term. Since
$G=\MeV{.4}\ll\eta+\kappa=\MeV{3.3}$, this is small compared to the
contribution of the second term. Evidently, this discussion of
$E^\ast_2$ applies analogously to the true Fock term without the term
proportional to $\eta$.

As seen from Fig.~\ref{fig1}, the RPA contribution to the symmetry
energy is in the pairing plus symmetry force model largely taken into
account by adding $\mu/2$ to the self-consistent energy $E_0$. This
corresponds to the approximation employed by Frauendorf and Sheikh, who
added $\mu/2$ to an isocranked Hartree-Fock-Bogolyubov
energy~\cite{FrSh}. As pointed out by these authors, doing so is
essentially equivalent to assigning the energy calculated for $\langle
T_\alpha\rangle=T+1/2$ or $\sqrt{T(T+1)}$ to a state with isospin $T$,
where $T_\alpha$ is some isospin coordinate. In fact, if $E_0(\langle
T_\alpha\rangle)$ is the calculated energy, we have
$E_0\left(\sqrt{T(T+1)}\;\right)\approx E_0(T+1/2)\approx
E_0(T)+E'_0(T)/2=E_0(T)+\mu/2$. Such procedures are common in ordinary
cranking calculations. See references in Ref.~\cite{FrSh}. In a context
of isocranking, Satula and Wyss applied the constraint $\langle
T_x\rangle=\sqrt{T(T+1)}$~\cite{SaWy01}. In the pairing plus symmetry
force model, the $T$-dependence of the RPA correlation energy is seen to
be taken into account by such recipes only in a crude approximation:
$E_2$ has a $T$-dependence additional to that of $\mu/2$. It is in fact
this additional $T$-dependence which makes $x>1$.

The presence of a strong isovector pair field is essential for getting
$x\approx1$. For $G=0$ the Hartree-Fock approximation is in fact exact,
and we have $E-E_{T=0}=(\eta T^2+\kappa T(T+1))/2=(\eta+\kappa)/2\times
T(T+x)$, where $x=\kappa/(\eta+\kappa)=.7$ for $\eta=\MeV{1.1}$. The
leading contribution of the pairing force in perturbation theory
$\langle-G\mathbf P^\dagger\cdot\mathbf P\rangle=G(T-3\Omega)/2$ adds a
linear term $GT/2$ to the symmetry energy, but as long as $G$ is below
the threshold of pair condensation $G\approx\eta/(2+\log(\Omega/2))$ it
is much smaller than the term $\eta T/2$ required to get $x=1$. A pair
gap $\Delta$ which is comparable to the single-nucleon level spacing
$\eta$ is thus necessary in order to make the quasiparticle energies and
the RPA frequencies except $\mu$ so independent of $T$ that $\mu/2$
becomes the dominant term in $E_2-E_{2,T=0}$.

In conclusion, the pairing plus symmetry force model with equidistant
single-nucleon levels introduced in Ref.~\cite{Ne} gives in the RPA for
$A=48$ a symmetry energy proportional to $T(T+x)$, where
$x\approx1.025$. The large iso\-vector pair field, whose strength is
inferred from the observed odd-even mass staggering, is essential for
this result. The RPA contribution to the symmetry energy is roughly
equal to $\mu/2$, where $\mu$ is the isocranking angular velocity, but
there are significant corrections to this approximation. It is in fact
these corrections which make $x>1$. The RPA contribution to the symmetry
energy cannot be simulated by adding to the Hartree-Fock-Bogolyubov
energy a term proportional to the isospin variance in the Bogolyubov
quasiparticle vacuum.

\bibliography{wigner5}

\end{document}